\documentclass[prb,twocolumn,amsmath,amssymb,floatfix,superscriptaddress,aps]{revtex4-2}

\usepackage{color}
\usepackage{graphicx}
\usepackage{hyperref}
\usepackage{braket}
\usepackage[capitalize]{cleveref}
\usepackage[normalem]{ulem}
\usepackage{bbm}
\usepackage{soul}

\renewcommand{\Im}{{\rm Im}}

\begin{document}
	
	\title{Analytic approach to thermoelectric transport in double quantum dots }
	
	\author{Nahual Sobrino}\email{nsobrino@ictp.it}
	\affiliation{Nano-Bio Spectroscopy Group and European Theoretical Spectroscopy Facility (ETSF), Departamento de Pol\'imeros y Materiales Avanzados: F\'isica, Qu\'imica y Tecnolog\'ia, Universidad del Pa\'is Vasco UPV/EHU, Avenida de Tolosa 72, E-20018 San Sebasti\'an, Spain}
	\affiliation{The Abdus Salam International Center for Theoretical Physics (ICTP), Strada Costiera 11, 34151 Trieste, Italy}

	\author{David Jacob}
	\affiliation{Nano-Bio Spectroscopy Group and European Theoretical Spectroscopy Facility (ETSF), Departamento de Pol\'imeros y Materiales Avanzados: F\'isica, Qu\'imica y Tecnolog\'ia, Universidad del Pa\'is Vasco UPV/EHU, Avenida de Tolosa 72, E-20018 San Sebasti\'an, Spain}
	\affiliation{IKERBASQUE, Basque Foundation for Science, Plaza Euskadi 5, E-48009 Bilbao, Spain}
    \affiliation{Departamento de F\'isica Aplicada, Universidad de Alicante, Campus de San Vicente del Raspeig, E-03690 Alicante, Spain}
	
	\author{Stefan Kurth}
	\affiliation{Nano-Bio Spectroscopy Group and European Theoretical Spectroscopy Facility (ETSF), Departamento de Pol\'imeros y Materiales Avanzados: F\'isica, Qu\'imica y Tecnolog\'ia, Universidad del Pa\'is Vasco UPV/EHU, Avenida de Tolosa 72, E-20018 San Sebasti\'an, Spain}
	\affiliation{IKERBASQUE, Basque Foundation for Science, Plaza Euskadi 5, E-48009 Bilbao, Spain}
	\affiliation{Donostia International Physics Center (DIPC), Paseo Manuel de
		Lardizabal 4, E-20018 San Sebasti\'{a}n, Spain}
	\date{\today}
	\raggedbottom
	
\begin{abstract}
	A recently proposed analytical solution for the equations of motion
	of the one-body Green function of the double quantum dot is extended to the
	out-of-equilibrium situation. By solving a linear system for the density
	correlators, not only the local occupations but also charge and heat currents
    as well as transport coefficients and the figure of merit are analytically
    derived in terms of system parameters and external driving forces. The emerging regions of stable occupation and finite currents are explained in terms of addition and removal energies, corresponding to the poles of the Green function. The analytical results are validated against the hierarchical equations of motion method, showing excellent agreement.
\end{abstract}

	\maketitle

\section{Introduction} 

The study of thermoelectric transport in nanoscale systems is a rapidly advancing field, driven by the exploration of quantum effects in low-dimensional systems, the need for efficient energy conversion technologies, and advancements in nanofabrication techniques, among others \cite{dresselhaus2007new, snyder2008complex, bell2008cooling, giazotto2006opportunities, zebarjadi2012perspectives, shakouri2011recent, vineis2010nanostructured, pichanusakorn2010nanostructured}. In this context, quantum dots (QDs) have gained significant research interest due to their nanoscale dimensions, discrete energy levels, and highly tunable electronic properties \cite{alivisatos1996perspectives, bayer2001coupling, kastner1993artificial, reed1988observation}. Double quantum dots (DQDs), in particular, offer additional control over electronic states through interdot coupling and Coulomb repulsion, which allows for the manipulation of charge and spin states, coherence effects, and enhanced interaction dynamics, making them promising candidates for thermoelectric applications \cite{vanderwiel2002electron, hanson2007spins, petta2005coherent, juergens2013thermoelectric, donsa2014double, zimbovskaya2020thermoelectric, pirot2022thermal1}.

While much research has traditionally focused on the equilibrium properties of DQDs \cite{you1999spectral, lamba2000transport, vzitko2010fano, georges1999electronic, busser2000transport}, these studies, although very useful and interesting, do not capture the full complexity of these systems under practical operating conditions. In real-world applications, devices often operate under nonequilibrium conditions, influenced by external driving forces such as thermal gradients and bias voltages. Understanding the behavior of DQDs in these out-of-equilibrium situations is crucial for optimizing their performance in thermoelectric devices and other nanoscale applications \cite{meir1991landauer, jauho1994time, haug2008quantum, datta1995electronic, di2013nonequilibrium, levy2013steady, sztenkiel2007electron, kuo2007tunneling, pohjola1997resonant, niu1995coherent, sun2002double, chi2006interdot, pirot2022thermal2, perez2023thermoelectric, cheng2021thermoelectric, tesser2022heat}.

Various theoretical techniques have been developed to solve non-equilibrium open quantum systems, including the Hierarchical Equations of Motion (HEOM), Numerical Renormalization Group (NRG), Quantum Master Equations (QME), and Dynamical Mean-Field Theory (DMFT) \cite{tanimura2020numerically, wilson1975renormalization, li2005quantum, georges1996dynamical, breuer2002theory, ryndyk2009green}. Among these, the equation of motion (EOM) method stands out as a powerful technique that provides a systematic approach to deriving the equations governing the dynamics of Green functions (GFs) in Hubbard and impurity models \cite{zubarev1960double, hubbard1963electron, hubbard1964electronII, hubbard1964electronIII, meir1992}. The EOM technique has successfully been applied to single and multiple QD systems, allowing for a numerical evaluation of the system's GF in equilibrium and non-equilibrium situations \cite{kang1995equation, van2010anderson, kang1998transport, swirkowicz2003nonequilibrium, sierra2016interactions, alomar2016coulomb, chang2008theory}. Given the complex structure of the equations, the solution is typically obtained through a self-consistent numerical procedure. Recently, an alternative fully analytical derivation of the EOM for the DQD in
the Coulomb blockade regime at thermal equilibrium has been proposed in Ref.~\cite{sobrino2024fully}. This approach provides explicit functional dependencies that are crucial for understanding the underlying physical mechanisms governing orbital occupations and spectra, both of which are essential for designing efficient nanoscale devices.

Here we extend the analytical derivations to the nonequilibrium
  situation and obtain expressions for key quantities such as orbital
  occupations, charge and heat currents, and transport coefficients in terms
  of the system parameters. 
These expressions will enable us to systematically study the influence of these parameters  and external driving forces on the thermoelectric characteristics of DQDs. 

The rest of the paper is organized as follows: In \cref{section_2}, we introduce the DQD Hamiltonian coupled to reservoirs, and derive the expressions for charge and heat currents using the EOM approach followed by the transport coefficients and the figure of merit. In \cref{section_4}, we present our results and validate them against HEOM numerical simulations. Finally, \cref{section_5} concludes the paper with a summary of our findings.
\section{Model and Currents from the Equations of Motion}
\label{section_2}
\subsection{Double Quantum Dot Hamiltonian}

We consider a parallel double quantum dot system attached to two electron reservoirs. The reservoirs are in local thermal equilibrium with temperatures $T_L$ and $T_R$, and chemical potentials $\mu_L$ and $\mu_R$, respectively. The Hamiltonian of the system is given by

\begin{align}
  \hat{\mathcal{H}}=&\hat{\mathcal{H}}_0+
  \sum_{i \alpha k \sigma}\epsilon_{\alpha k i}\hat c^{\dagger}_{\alpha k i\sigma}\hat c_{\alpha k i\sigma}\nonumber\\
  &+\sum_{i \alpha k \sigma }\left(V_{\alpha k i}\hat c^{\dagger}_{\alpha k i\sigma}\hat d_{i \sigma}+\text{H.c.}\right)\;,
	\label{eq_H}
\end{align}
where 
\begin{align}
	\hat{\mathcal{H}}_0 = \sum_i v_{i}\hat n_{i}+\sum_i U_{i}\hat n_{i\sigma}\hat n_{i\bar\sigma}+U_{12}\hat n_{1}\hat n_{2}
	\label{eq_H0}
\end{align}
describes the isolated double dot. Here, $\hat d^{\dagger}_{i\sigma}$ ($\hat d_{ i\sigma}$) is the creation (anhilation) operator for an electron with spin $\sigma$
   on dot $i$.  $\hat c^{\dagger}_{\alpha k i\sigma}$ ($\hat c_{\alpha k i\sigma}$)  is the creation (anhilation) operator for an electron with spin $\sigma$ in state $k$ of lead $\alpha$ coupled to the site $i$. Moreover, $v_i$ and
  $U_{i}$ are the on-site energy and the intra-Coulomb repulsion of dot $i$, respectively, while $U_{12}$ is the inter-Coulomb repulsion between the two dots. The total and spin-resolved density operators are defined as $\hat n_{i}=\sum_{\sigma=\uparrow,\downarrow}\hat n_{i\sigma}$ and $\hat n_{i\sigma}=\hat d^{\dagger}_{i\sigma}\hat d_{i\sigma}$, respectively. The last two terms in \cref{eq_H} describe the single-particle eigenstates of the isolated reservoirs and the tunneling between the dots and the two reservoirs $\alpha=L,R$, with $V_{\alpha k i}$ as the coupling parameter. In the following, we work in the wide band limit, i.e., the reservoirs are featureless leads  at chemical potential $\mu_{\alpha}$ and the hybridization function or embedding
  self-energy $\Delta_{i}^{\alpha}\equiv\sum_{k}|V_{\alpha k i}|^2/(\omega-\epsilon_{\alpha k i})$
  becomes energy-independent.
  Additionally, we focus on the  situation where 
  both dots couple to the leads with the same coupling strength
    $\gamma_\alpha$ so that $\Delta_{i}^\alpha=-i\gamma_\alpha/4$.
The system can be brought out of equilibrium by imposing a thermal gradient $\Delta T = T_L - T_R$ and/or an external DC bias $V=\mu_L-\mu_R$ across the junction. It is assumed that, in the long-time limit, these perturbations result in a steady-state with an electrical current ($I$) and energy and heat currents ($W$ and $Q$, respectively). 

In the following, we use the sign convention that currents flowing into the central region are positive. Due to the conservation of charge and energy, the steady-state electrical/energy current entering from the left lead equals the steady-state electrical/energy current exiting through the right lead, i.e., $I \equiv I_L = -I_R$ (electrical current), $W \equiv W_L = -W_R$ (energy current), and $Q \equiv Q_L = P- Q_R$ (heat current) where $P=-IV$ is the electrical power. Furthermore the energy current of lead $\alpha$ is related to the charge and heat currents as $W_{\alpha}=Q_{\alpha}+\mu_{\alpha}I_{\alpha}$.

\begin{figure}
	\centering
	\includegraphics[width=0.7\linewidth]{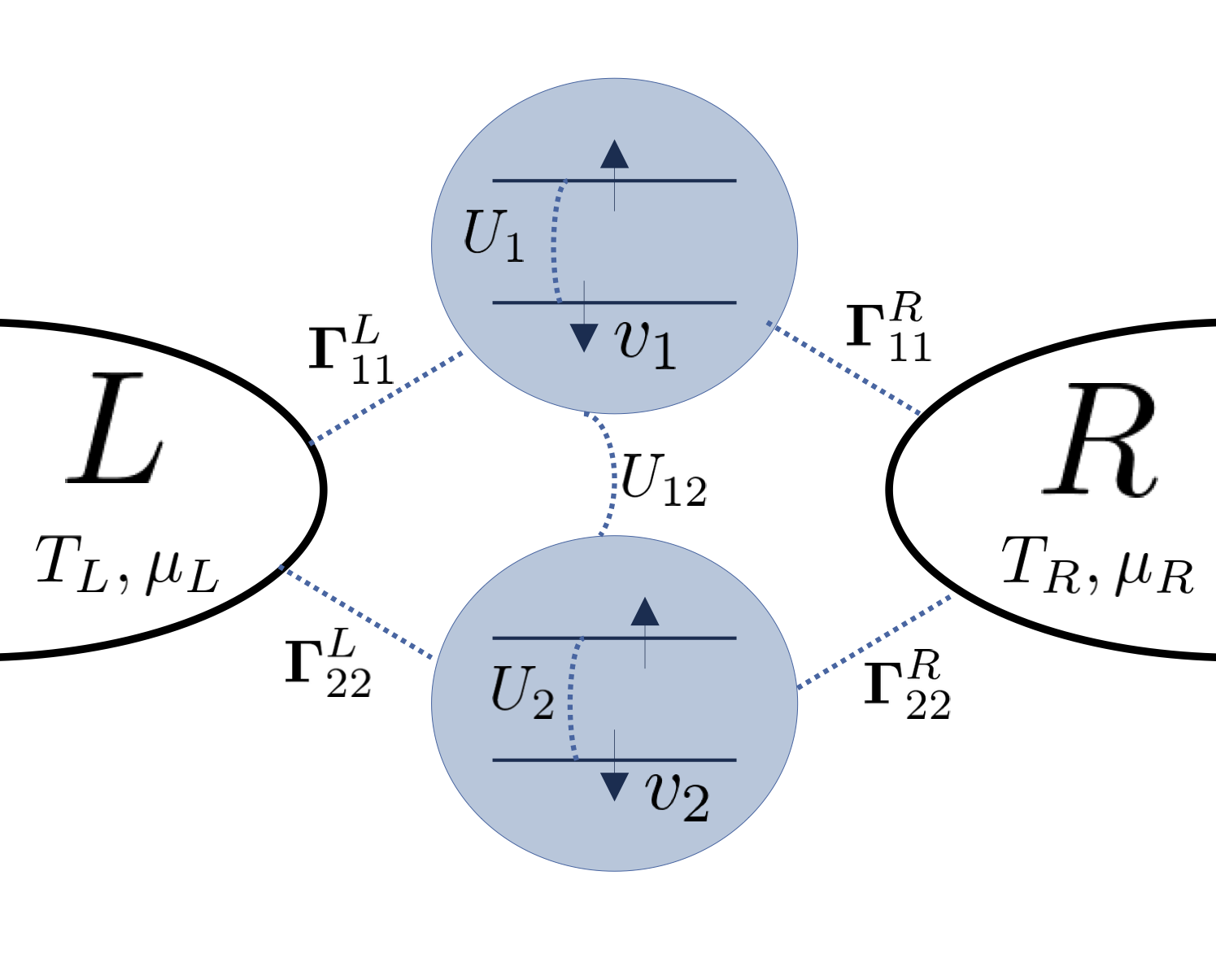}
	\caption{ Schematic representation of the transport setup for the DQD. The leads are coupled to the dots with the same coupling strength  $\mathbf{\Gamma}_{ij}^\alpha=\delta_{ij}\,\gamma_{\alpha}/2$.}
	\label{fig:figure1}
\end{figure}

\subsection{Charge and Heat Currents from the Equation of Motion Approach}

The charge and heat currents through an interacting region attached to two non-interacting electronic reservoirs can be derived using the non-equilibrium Keldysh formalism \cite{ludovico2014dynamical,meir1992}  ($\int \equiv \int\frac{d\omega}{2\pi}$ in the following):
\begin{subequations}
  \begin{align}
    I =&\frac{i}{2}\int \left[ \text{Tr}\left\{\left(f_L(\omega)\mathbf{\Gamma}^L-f_R(\omega)\mathbf{\Gamma}^R\right)(\mathbf{G}^r(\omega)-\mathbf{G}^a(\omega))\right\}\right.\nonumber\\
    &\left.+\text{Tr}\left\{ (\mathbf{\Gamma}^L-\mathbf{\Gamma}^R)\mathbf{G}^{<}(\omega)\right\}\right]\;,\\
    Q  =&\frac{i}{2}\int (\omega-V_{L})\left[ \text{Tr}\left\{\left(f_L(\omega)\mathbf{\Gamma}^L(\omega)-f_R(\omega)\mathbf{\Gamma}^R\right)\right.\right.\nonumber\\
    &\left.\left.\times(\mathbf{G}^r(\omega)-\mathbf{G}^a(\omega))\right\}+\text{Tr}\left\{ (\mathbf{\Gamma}^L-\mathbf{\Gamma}^R)\mathbf G^{<}(\omega)\right\}\right] \;,
    \label{eq_I_Q_MW}
  \end{align}
\end{subequations}
where $f_{\alpha}(\omega)=[1+e^{\frac{\omega-\mu_{\alpha}}{T_\alpha}}]^{-1}$ is the Fermi distribution of lead $\alpha=L,R$. The retarded $\mathbf{G}^r(\omega)$, advanced $\mathbf{G}^a(\omega)$, and lesser $\mathbf{G}^<(\omega)$ Green functions (GF) are the Fourier transforms of the matrices with elements $G_{ij}^r(t)=-i \theta(t) \left\langle \left\{ \hat d_i(t), \hat d_j^\dagger(0) \right\} \right\rangle$, $G_{ij}^a(t)=i \theta(-t) \left\langle \left\{ \hat d_i(t), \hat d_j^\dagger(0) \right\} \right\rangle$, and $G_{ij}^<(t)=i \left\langle \hat d_i^\dagger(0) \hat d_j(t) \right\rangle$, respectively, with $\theta(t)$ as the Heaviside step function, and the brackets $\braket{\dots}$ indicating thermal average.
$\mathbf{\Gamma}^\alpha$ are the so-called coupling matrices
  describing the coupling coupling of the DQD to each of the leads, and
  defined as the anti-hermitian part of the embedding self-energies, defined above.
  Here the embedding-self energies and thus the coupling matrices are diagonal in the dot indices $i$,
  and hence $\mathbf{\Gamma}^\alpha_{ij}=-2\,\Im\,\Delta_i^\alpha\,\delta_{ij}$. In the symmetrically coupled situation, $\mathbf{\Gamma}_{ij}^{\alpha}= \delta_{ij}\,{\gamma_{\alpha}/2}$, considered here,
the currents through the DQD system of \cref{eq_H} simplify to
\begin{subequations}
  \begin{align}
    I=& - \frac{\gamma}{2} \sum_i \int \left(f_L(\omega)-f_R(\omega)\right)\text{Im}(G^{r}_i(\omega))\;,\\
    Q=& - \frac{\gamma}{2} \sum_i \int (\omega-V_L)\left(f_L(\omega)-f_R(\omega)\right)\text{Im}(G^{r}_i(\omega)) \; .
  \label{eq_I_Q}
  \end{align}
\end{subequations}

The equation-of-motion (EOM) approach allows for the calculation of the
one-particle GF in terms of higher order GFs whose EOM in turn generates yet higher order GFs. In order for the EOMs
to be practically useful, this hierarchy needs to be truncated. Here we
employ the (approximate) truncation scheme of Ref.~\cite{sobrino2024fully}
which may be generalized with the same arguments to the out-of-equilibrium
steady state situation. Our truncation scheme may be rationalized in three
differen ways \cite{sobrino2024fully}, all leading to the same
approximation. One of these rationalizations neglects certain higher-order
GFs but the same result can also be achieved by simply broadening all poles
(see below) of all GFs by the coupling parameter
$\gamma = \gamma_L + \gamma_R$.
Finally, the assumption that the local density operator and the Hamiltonian
commute, i.e., $[ \hat n_{i\sigma},\mathcal{H}]\approx 0$ also leads to the
same truncation scheme. 
This approximation is accurate in the Coulomb blockade regime, when the
temperature $kT$ is large compared to the broadening $-\text{Im}\Delta_i$ by
the reservoirs. Furthermore, we approximate the correlators in the
out-of-equilibrium situation by the phenomenological expression \cite{bulka2004electronic,kuo2011theory,chang2008theory,t2012effects}
\begin{align}
	&\braket{n_{i'\sigma'}\dots n_{i''\sigma''}n_{i'''\sigma'''}}=-\int \tilde f(\omega)\nonumber\\
	&\times\text{Im}\left(\braket{\braket{n_{i'\sigma'}\dots n_{i''\sigma''}d_{i'''\sigma'''}:d^{\dagger}_{i'''\sigma'''}}}\right)\;,
	\label{eq_exp_values}
\end{align}
where $\tilde f (\omega)= \frac{1}{2}(f_L(\omega)+f_R(\omega))$. This is a
straightforward generalization of the corresponding (exact) equilibrium
expression (see \cite{sobrino2024fully}) to which it correctly reduces
  in thermal equilibrium. In particular, the one-body correlators correspond
to the local occupation 
\begin{align}
	\braket{n_{i\sigma}}=-\int \tilde f(\omega)\text{Im}\left(G_{i\sigma}^{r}(\omega)\right)\;.
	\label{eq_exp_values1}
\end{align}
\cref{eq_exp_values} is an approximation in the out-of-equilibrium situation.

In Ref.~\cite{sobrino2024fully}, we derived an analytical expression for the single-particle GF at equilibrium for the DQD system given by \cref{eq_H}.  With the assumption \cref{eq_exp_values} for the non-equilibrium correlators this derivation can be transferred one-to-one to the non-equilibrium situation considered here, resulting in an expression for the GF solely in terms of the occupations $\langle\hat{n}_i\rangle$ of each QD $i$ and the electron addition and removal energies
\begin{align}
	&G_{i\sigma}^r(\omega)=\sum_{j=1}^{6}\frac{r_{i,j}}{\omega-p_{i,j}+i\frac{\gamma}{2}}\;,
	\label{eq_GF0}
\end{align}
where the poles $p_{i,j}$ correspond to the addition and removal energies
	\begin{align}
		&p_{i,1}=v_{i} \;,\quad & &p_{i,4}=v_{i}+U_{i}+2 U_{12}\;,\nonumber \\
		&p_{i,2}=v_{i}+U_{i}\;, \quad & &p_{i,5}=v_{i}+U_{12}\;,\nonumber \\
		&p_{i,3}=v_{i}+U_{i}+U_{12}\;, \quad & &p_{i,6}=v_{i}+2 U_{12} \;,
			\label{eq_pi}
	\end{align}
and the residues $r_{i,j}$  are linear combinations of the local occupations (see Eq.~(23) of Ref.~\cite{sobrino2024fully}), but computed via the non-equilibrium expression \cref{eq_exp_values1} instead of the equilibrium one.
The residues and the densities expressions analytically derived in Ref.~\cite{sobrino2024fully} depend on a function $\phi(p)$ that corresponds to the integral of a single pole in the complex plane multiplied by the Fermi distribution. In the out of equilibrium situation, the explicit dependence on the local chemical potentials and temperatures of each of the leads in \cref{eq_exp_values}, make the results completely analogous to the equilibrium situation with the modified function $\phi(p)$ which now reads:
\begin{align}
\phi(p)&=\int  \tilde f(\omega)\frac{\gamma}{(\omega-p)^{2}+\frac{\gamma^{2}}{4}}=\frac{1}{2}-\sum_{\alpha}\frac{1}{2\pi}\text{Im}[\psi\left(z_{\alpha} \right)]\;,
\label{eq_int_digamma}
\end{align}
where $z_\alpha = \frac{1}{2}+\frac{\gamma/2+i(p-V_\alpha)}{2\pi T_\alpha}$ for $\alpha=L,R$, and   $\psi(z)=\frac{d \log(\Gamma(z))}{dz}$ is the digamma function with general complex argument $z$, and $\Gamma(z)$ is the gamma function. 


Given the structure of the GF as a sum of single poles in the complex plane, the current integrals \cref{eq_I_Q}  can also be computed analytically \cite{sobrino2021thermoelectric}, leading to the expressions
\begin{subequations}
\begin{align}
	I=& -\frac{\gamma}{2\pi} \sum_{i,j,\alpha}s_\alpha r_{i,j}\text{Im}[\psi(z^{\alpha}_{i,j})]\;,\\
	Q=& \frac{\gamma}{2\pi} \sum_{i,j,\alpha}s_\alpha r_{i,j}\left[ \frac{\gamma}{2}\text{Re}[\psi(z^{\alpha}_{i,j})]-(p_{i,j}-\frac{V}{2})\text{Im}[\psi(z^{\alpha}_{i,j})]\right]\nonumber\\
	&+\frac{\gamma^2}{2\pi}\log\left(\frac{T_L}{T_R}\right)\;,
\end{align}
\label{eq_currents_digamma}
\end{subequations}
with $s_\alpha=\pm1$ for $\alpha=L,R$, and $z^{\alpha}_{i,j}=\frac{1}{2}+\frac{\frac{\gamma}{2}+i(p_{i,j}-V_\alpha)}{2\pi T_\alpha}$. 

\subsection{Linear Transport coefficients}
\label{section_3}
The linear response relationship between the currents and the external potentials reads
\begin{align}
\begin{pmatrix}
	I \\
	Q
\end{pmatrix}
=
\begin{pmatrix}
	L_{11} & L_{12}\\
	L_{21} & L_{22}
\end{pmatrix}
\begin{pmatrix}
	V/T \\
	\Delta T/T^2
\end{pmatrix}\;,
\label{eq_LR_IQ}
\end{align}
with $L_{12}=L_{21}$ from Onsager's relation \cite{onsager1931reciprocal}. 
Taking the derivatives in \cref{eq_currents_digamma} with respect to the external potentials, we can analytically  derive the matrix elements $L_{ij}$ of the conductance matrix as
\begin{subequations}
\begin{align}
	L_{11}&= T\left. \frac{\partial I}{\partial V} \right|_{\substack{V=0 \\ \Delta T=0}}=\frac{\gamma}{4\pi^2}\sum_{i,j} r_{i,j}\text{Im}[i\psi'(z_{i,j})] \;,\\
	L_{12}&=T^2 \left. \frac{\partial I}{\partial \Delta T} \right|_{\substack{V=0 \\ \Delta T=0}}=\frac{\gamma}{4\pi^2}\sum_{i,j} r_{i,j}\text{Im}[l_{i,j}^0\psi'(z_{i,j})] \;,\\
	L_{22}&=T^2 \left. \frac{\partial Q}{\partial \Delta T} \right|_{\substack{V=0 \\ \Delta T=0}}=\frac{\gamma}{4\pi^2}\sum_{i,j}r_{i,j}\left[ p_{i,j}\text{Im}[l_{i,j}^0\psi'(z_{i,j})]\right.\nonumber\\
	& \left.-\frac{\gamma}{2}\text{Re}[l_{i,j}^0\psi'(z_{i,j})]  \right]+\frac{\gamma^{2}T}{2\pi}\;,
\end{align}
\end{subequations}
where $l_{i,j}^0=\frac{\gamma}{2}+ip_{i,j}$, $z_{i,j}=\frac{1}{2}+\frac{l_{i,j}^0}{2\pi T}$, and $\psi'(z)=\frac{d^2 \log(\Gamma(z))}{dz^2}$ is the trigamma function with complex argument $z$.\cite{abramowitz1968handbook}
The linear transport coefficients, i.e., the electrical conductance $\mathcal{G}$, the Seebeck coefficient $\mathcal{S}$ and the thermal conductance $\kappa$, can be expressed in terms of the conductance matrix elements $L_{ij}$ as
\begin{subequations}
\begin{align}
	\mathcal{G} &= \left. \frac{\partial I}{\partial V} \right|_{\substack{V=0 \\ \Delta T=0}}=\frac{L_{11}}{T}\;,\\
    	\mathcal{S} &= -\left. \frac{\partial V}{\partial \Delta T} \right|_{\substack{I=0 \\ Q=0}}=\frac{L_{12}}{T L_{11}}\;,\\
	\kappa &= \left. \frac{\partial Q}{\partial \Delta T} \right|_{\substack{I=0 \\ Q=0}}=\frac{1}{T^{2}}\left(L_{22}-\frac{L_{12}^2}{L_{11}}\right)\;.
\end{align}
\label{eq_LTC}
\end{subequations}

From \cref{eq_LTC} one can compute the dimensionless figure of merit $\text{ZT}$, a key parameter in evaluating the efficiency of thermoelectric materials. It provides a measure of the system's ability to convert heat into electricity
\begin{align}
\text{ZT} = \frac{	\mathcal{S}^2 	\mathcal{G} T}{\kappa}.
\end{align}
For a system to have efficient thermoelectric properties, we desire $\text{ZT}$ to be as high as possible, which implies a high Seebeck coefficient, high electrical conductance, and low thermal conductance.  The thermal conductance, $\kappa$, consists of the electronic part, $\kappa_e$, and the phonon part, $\kappa_{\text{ph}}$. In our analysis, the phonon contribution, which arises from phonon transport through the device, has been neglected. Therefore, the $\text{ZT}$ values calculated by excluding the phonon thermal conductivity represent the upper limits for each set of parameters.

\section{Results}
\label{section_4}
\begin{figure*}
\includegraphics[width=\linewidth]{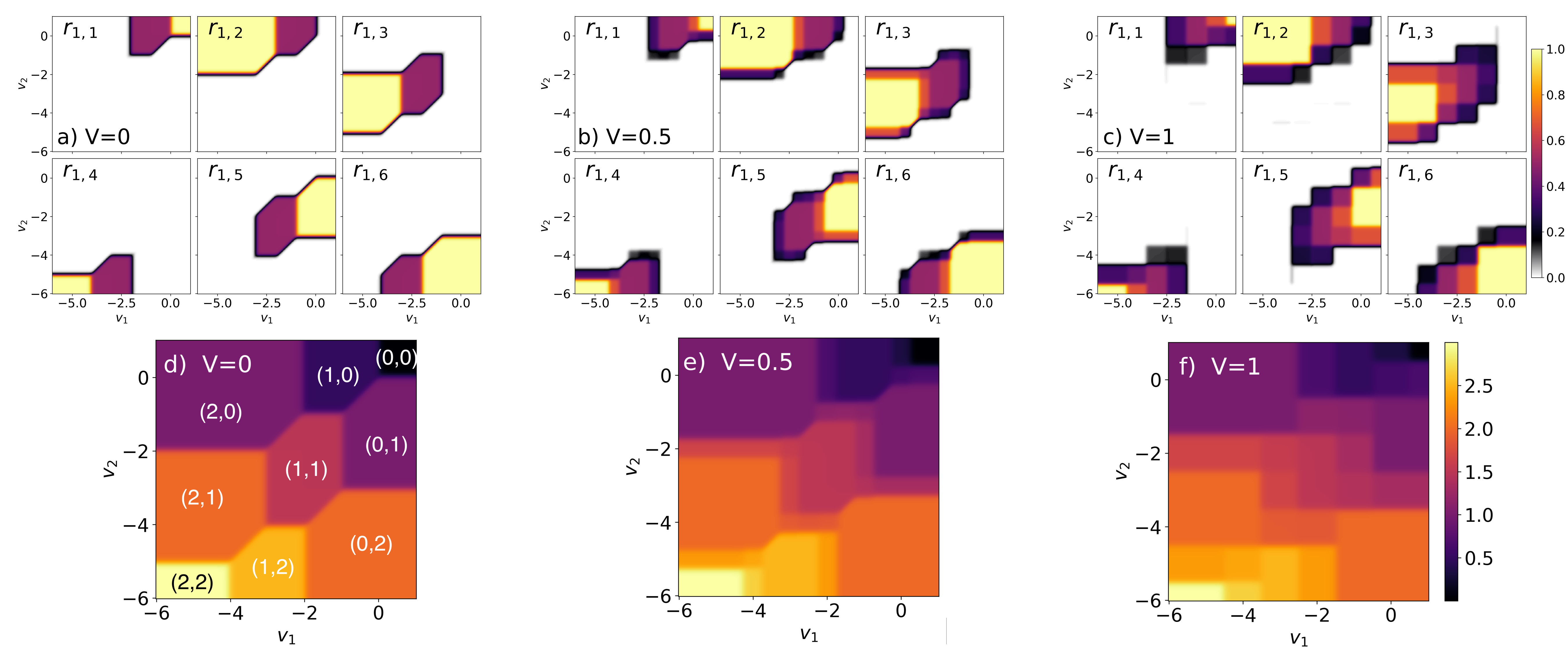}
\caption{\label{fig2_res_SD}
  Residues $r_{1,j}$ of the Green function (top row) and stability diagrams (bottom row) as functions of gate potentials $v_i$ for different
  symmetrically applied bias voltages $V$. Panels (a) and (d) correspond to $V=0$, panels (b) and (e) correspond to $V=0.5$, and panels (c) and (f) correspond to $V=1$. In panel (d) we also indicate the occupations $(n_1,n_2)$ for different regions of the stability diagram. The parameters used are $U_1=2$, $U_2=3$, $T=0.05$, and $\gamma \to 0$. Energies in units of $U_{12}$.
}
\end{figure*}
	In this section, we implement the analytical EOM method to calculate densities, currents, transport coefficients and the figure of merit of the DQD across various regimes determined by the interaction parameters (see  Ref.~\cite{sobrino2020exchange}). We validate our findings by comparing them with the numerical hierarchical equations of motion (HEOM) approach \cite{tanimura2020numerically}. A Python implementation of our EOM method for the DQD is available on GitHub\cite{github_DQD}.

	In \cref{fig2_res_SD}, we show the residues $r_{1,j}$ of the GF (panels (a-c)) given by Eq.~(23) of Ref.~\cite{sobrino2024fully} evaluated with $\phi(p)$ from \cref{eq_int_digamma} and the stability diagrams (panels (d-f)) of the DQD system under varying gate voltages $v_1$ and $v_2$, and for different applied bias voltages between the electrodes $V$ for interaction parameters $U_1=2, U_2=3, U_{12}=1$. The stability diagrams represent regions in the $v_1-v_2$ plane where the occupation numbers are stable, i.e., they do not fluctuate in the limit of low temperature and weak coupling to the reservoirs.
        The lines separating regions in the stability diagrams correspond to parameters which lead to degenerate ground states. In the low temperature and low coupling limit and at $V=0$
        (\cref{fig2_res_SD} left), each residue has a finite contribution in a maximum of two different regions of the stability diagram. For fixed $(v_1, v_2)$, either only one non-vanishing residue with value 1 exists, or there exist two non-vanishing residues both with value 0.5 (in the limit of low temperature and weak coupling). This implies that the local spectral function on dot 1 has either one or two poles. At $V=0$, the occupations in the different regions of the stability diagram are stable at integer values.
	
        When a finite symmetric bias voltage is applied between the electrodes, the vertical and horizontal lines separating the different regions of the stability diagram in the $(v_1, v_2)$ plane split, defining new regions with fractional occupations in the stability diagram.
This splitting can be understood by the local bias dependencies in
  the two Fermi functions $\tilde{f}$ of \cref{eq_int_digamma}. In the stripe
  region, a given pole with finite residue in this region (see top panels of
  \cref{fig2_res_SD}), will fall between the local chemical potentials
  $\mu+V_{\alpha}$ ($\alpha = L,R$) entering $\tilde{f}$. Alternatively, one
  can understand the formation of the stripe regions in the finite-bias
  stability diagrams by performing a variable substitution
  $\omega'=\omega-V_{\alpha}$ separately for each of the Fermi functions
  contributing to $\tilde{f}$ of \cref{eq_int_digamma}. Then the contributions
  of the poles of the GF of site $i$ will be shifted to an effective gate
  level $v_i'=v_i-V_{\alpha}$.
The width of these new regions exactly corresponds to the value of the applied voltage $V=V_L-V_R$, 
and they are centered around lines separating regions of the $V=0$
  stability diagram. 
In the case of an asymmetric bias, the new
stripe regions are shifted asymmetrically with respect to the
degeneracy lines  (not shown).
	
  In  (\cref{fig2_res_SD}(b) ), where $V=0.5 = 2V_L = -2V_R$, the values of the residues along the stripe regions is either 0.75 or 0.25. In the corresponding stability diagram, the stripes correspond to stable regions of
  non-integer local occupation of one of the sites with values 2/3 or 4/3. This value can be deduced from the zero-bias degeneracy line from which the new stripe region originates. If the two regions separated by this degeneracy line (at equilibrium) have occupancy in one of the sites $n_i$ equal to 0 and 1, then the finite stripe region will have occupation $n_i=2/3$. On the other hand, if the two differing occupations for $n_i$ are 1 and 2 then the stripe region will have occupation $n_i=4/3$. The values 2/3 and 4/3 for the fractional occupations at finite bias also appear in the single-impurity Anderson model at finite bias \cite{StefanucciKurth:15}.  As the bias is increased further to $V=1$ (\cref{fig2_res_SD}(c) ), the new width of the stripe regions of both the residues and the stability diagram increase according to the bias value. When the  new vertical and horizontal stripe regions meet at a triple degenerate point, further new square regions are created in both the residues and the stability diagram. This results from the increased splitting and mixing of energy levels due to the higher bias voltage, leading to more complex patterns of electron occupation.
	
\begin{figure}
\centering
\includegraphics[width=1\linewidth]{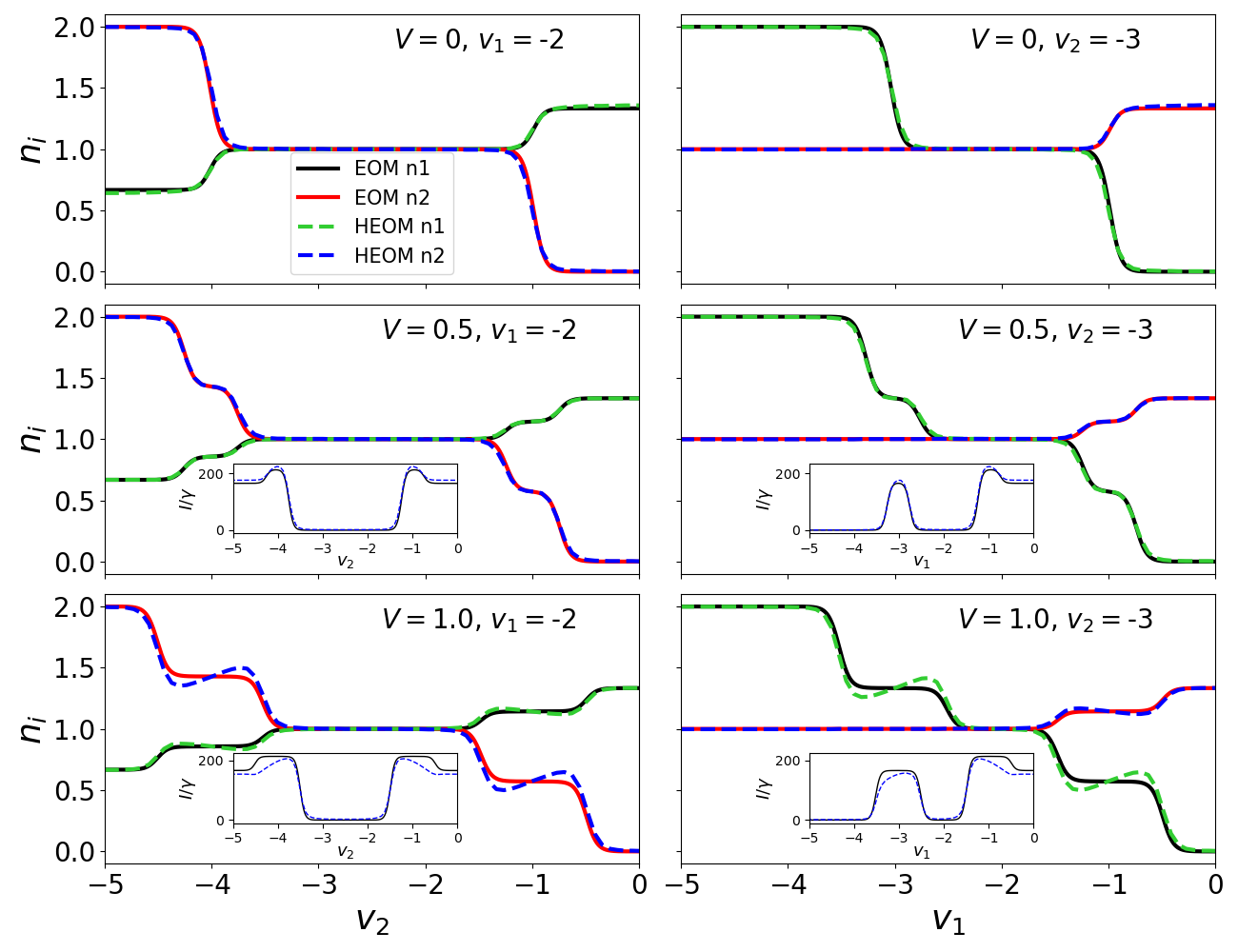}
\caption{Local occupations $n_i$ as functions of gate levels $v_2$ (left panels) and $v_1$ (right panels) for different bias voltages $V$. The insets show a comparison between the EOM (solid black) and the HEOM (dashed blue) charge currents. The interactions and parameters considered are $U_1=2$, $U_{2}=3$, $T=0.05$, and $\gamma=0.01$. Energies in units of $U_{12}$.}
\label{fig3_ni}
\end{figure}
        
In \cref{fig3_ni}, the local occupations $n_i$ and charge current $I$ (insets) are shown for specific gate potentials, corresponding to vertical and horizontal line cuts of the stability diagrams shown in \cref{fig2_res_SD}. These plots provide a detailed examination of how the occupations change with varying bias voltages at fixed interaction values. The left panels of \cref{fig3_ni} show the occupations $n_i$ as function of the gate voltage $v_2$ at $v_1 = -2$, while the right panels present the occupations as functions of $v_1$ at $v_2 = -3$. At $V = 0$ and fixed $v_1 = -2$, the occupations evolve as $v_2$ varies, with the main structure featuring three plateaus and two step transitions at $v_2 = -4$ and $v_2 = -1$. The non-integer values of $n_1$ correspond to the degenerate energy lines separating the regions in the stability diagram of \cref{fig2_res_SD}(d).  Similarly, at fixed $v_2 = -3$, $n_1$ evolves with steps at $v_1 = -3$ and $v_1 = -1$. In contrast, $n_2$ does not exhibit a step at $v_1 = -3$, as this transition corresponds to the ground state change from $(2,1)$ to $(1,1)$ in the low temperature limit. For $v_1 > -1$, the second dot is along a degenerate line, resulting in non-integer $n_2$. When a finite bias voltage of $V = 0.5$ is applied (central panels), an additional plateau appears between
the other plateaus, indicating the influence of the bias on the poles. The width of this new feature exactly corresponds to the applied bias, with an occupation corresponding to the value of half the occupation at the center of the step in the $V=0$ situation.  When the bias voltage increases to $V = 1$ (lower panels), the additional plateaus in occupations further broaden, corresponding to the new vertical and horizontal regions in the stability diagram (see \cref{fig2_res_SD}(f)). At relatively small bias, the analytical occupations and the charge current obtained from the EOM approach match exactly with the numerical results obtained from the HEOM method for tier level $L=3$. The only noticeable difference occurs at large bias ($V = 1$), where the new regions created by the finite bias are flat and stable in the analytical EOM results, while the HEOM predicts non-monotonous behavior
of the densities for varying local potentials $v_i$. In these regions, the HEOM local occupations show increases (or decreases) before continuing to decrease (or increase) again. The corresponding charge current shows a flatter behavior in the EOM results than in the HEOM, where the local plateaus of high currents is smoother.

\begin{figure*}
\centering
\includegraphics[width=0.85\linewidth]{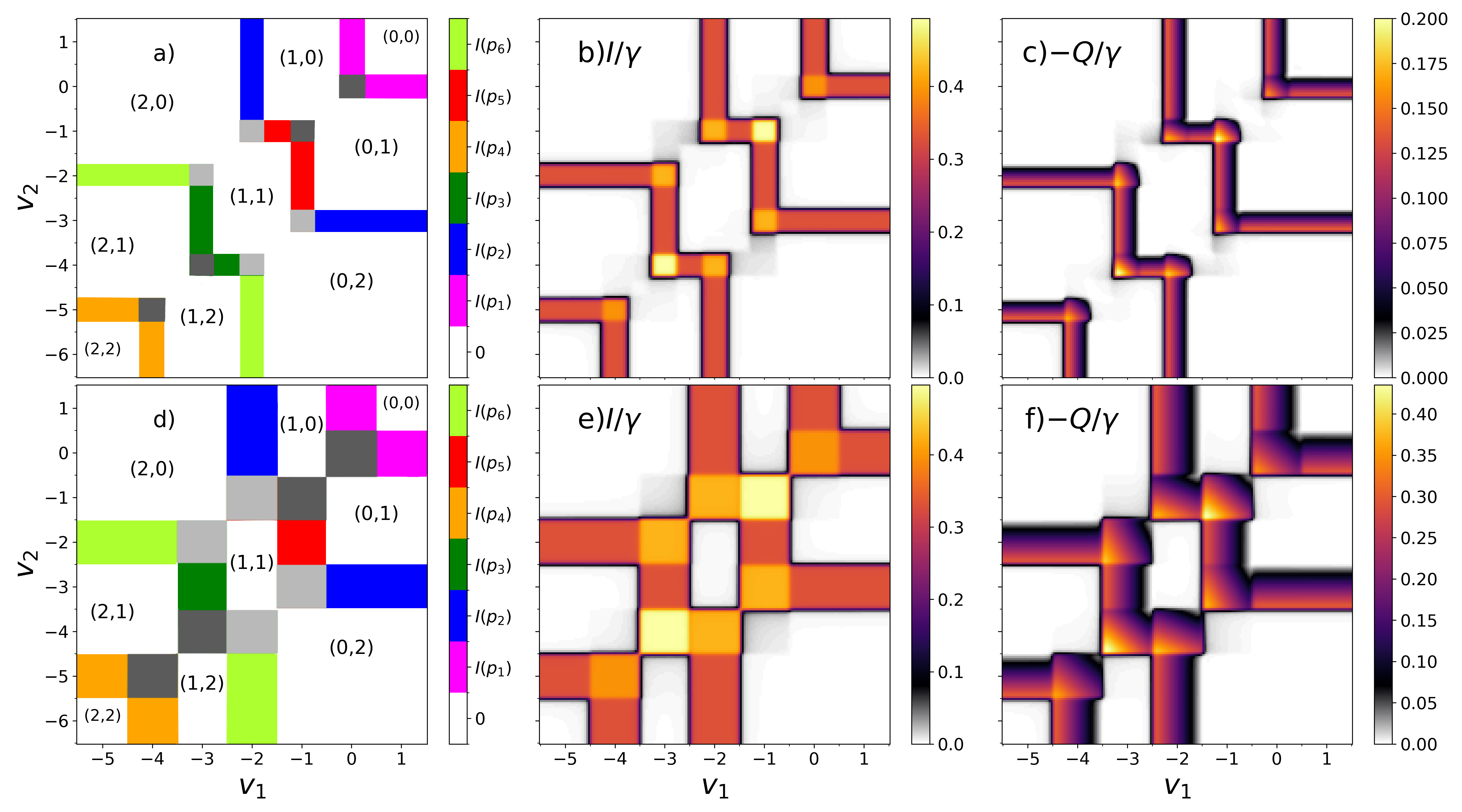}
\caption{ Charge and heat currents as functions of gate potentials $v_i$ for $V=0.5$ (top panels) and $V=1$ (bottom panels). The left panels show the contributions from the poles of the spectral function (Eq.~(23) of Ref.~\cite{sobrino2024fully} in the regions of finite currents. The other
  parameter and interaction values are $U_1=2$, $U_2=3$, $T=0.02$ and $\gamma=0.01$. Energies in units of $U_{12}$. }
\label{fig4_IQ_v1v2}
\end{figure*}

\begin{figure}
\centering
\includegraphics[width=1\linewidth]{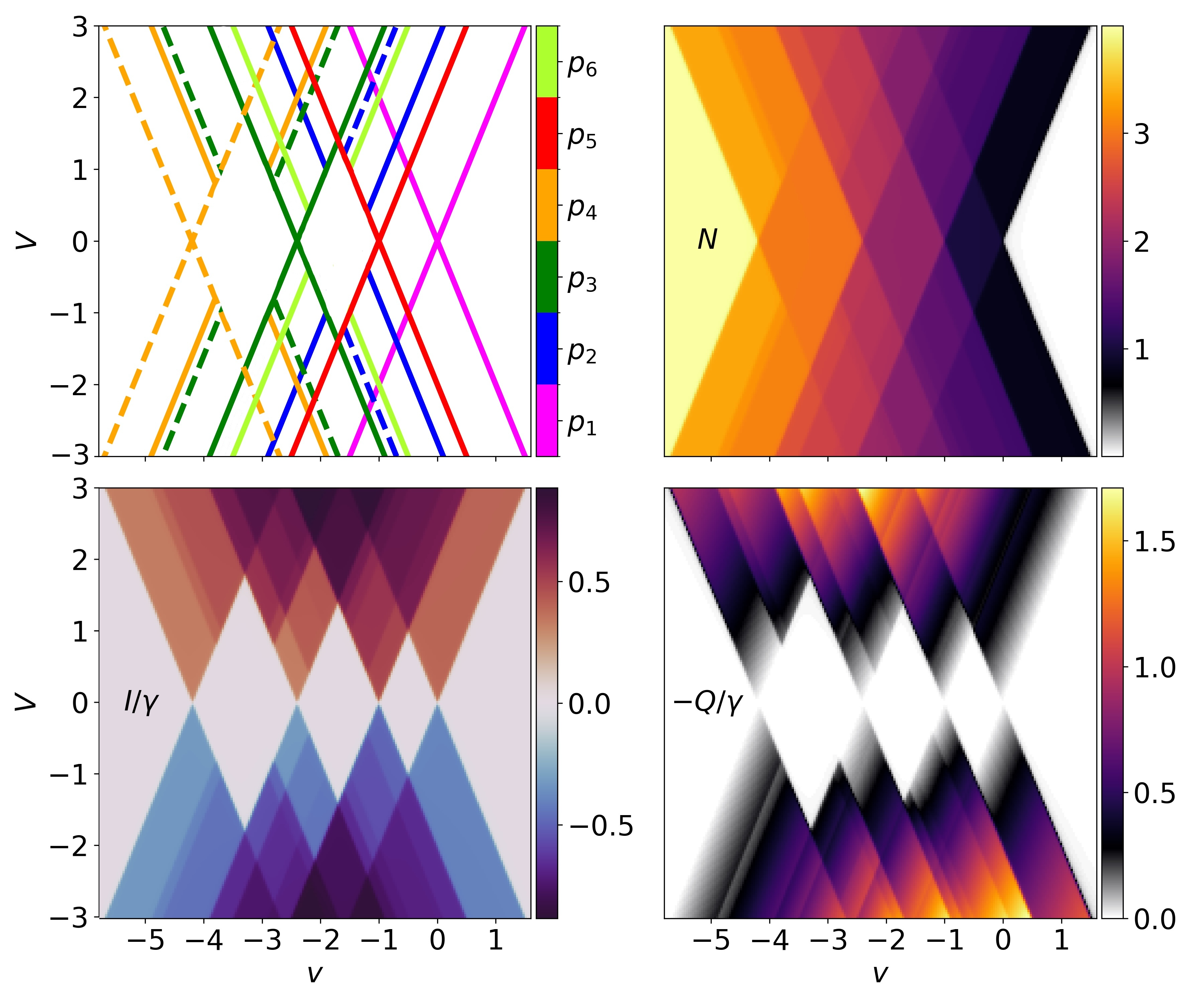}
\caption{
 Top left panel: positions of the poles with significant non-vanishing
    residues corresponding to solutions of $p_{i,j}-V_{\alpha}=0$ with
    $V_L=-V_R=V/2$ as function of gate $v=v_1=v_2$ and bias $V$. The lines with
    positive (negative) slopes correspond to $V_L$ ($V_R$), while the solid
    (dashed) lines correspond to poles related to site two (one). The other
    panels show the total occupation $N= n_1 + n_2$ (top right), the charge
    (bottom left) and the heat currents (bottom right). Here the parameters
    of the model are $U_1=2.2$, $U_2=1.4$, as well as $T=\gamma=0.01$.
    Energies in units of $U_{12}$.
}
\label{fig5_IQ_vV}
\end{figure}

\cref{fig4_IQ_v1v2} shows the charge and heat currents as functions of the gate voltages $(v_1,v_2)$ for different values of the bias $V$. The occurence of regions of finite charge and heat currents can be understood in terms of the poles of the GFs. As explained before, the finite bias contributions in the Fermi functions can be shifted to the poles of the spectral function, giving rise to effective gate levels. Since both the charge and heat current \cref{eq_I_Q} depend on the difference between the Fermi functions, the resulting contributions of each pole
cancel each other 
except for the stripe regions.  In the left panels of \cref{fig4_IQ_v1v2} we
show in different colors the contribution of each of the poles defined in
\cref{eq_pi} due to the different shift by the left and right bias.
These regions exactly correspond to the stripes that appear at finite bias in the stability diagrams. In particular, the $I(p_j)$ correspond to the finite current generated due the contribution of $p_{1,j}$ and $p_{2,j}$ (in the following we use $p_j$ as a collective variable which contains contributions both from $p_{1,j}$ and $p_{2,j}$)
\begin{align}
I(p_j)=& -\frac{\gamma}{2\pi} \sum_{i,\alpha}s_\alpha r_{i,j}\text{Im}[\psi(z^{\alpha}_{i,j})].
\end{align}
With this definition, the total current is $I=\sum_j I(p_j)$. In particular, the poles related to site one (two) will generate the stripe regions in the vertical (horizontal) directions in the $(v_1,v_2)$ plane. For instance, the stripe related to the pole $p_{2,2}=v_2+U_2$ (horizontal blue line in the left panels of \cref{fig4_IQ_v1v2}) will separate the regions of local occupations $(0,1)$ and $(0,2)$, since it exactly corresponds to the energy difference between the respective states.
The dark (light) gray squared regions have contributions from both adjacent equal (different) poles. In the central and right panels of \cref{fig4_IQ_v1v2} the charge  and heat currents are shown for $V=0.5$ (top panels) and $V=1$ (bottom panels). The  charge current
is essentially constant along the stripe regions, with a higher value in the squares where two of the finite pole contributions meet ($p_2$ and $p_5$ or $p_3$ and $p_6$). In particular, the two dark gray regions adjacent to the region with occupations $n_i=1$, {\em both} local occupations take one of the non-integer values of 3/4 and 5/4, and the charge current reaches its maximum value for that given bias. On the other hand, the heat current varies along the perpendicular direction of the stripes, increasing its value as the gate related to the other site is increased. The small finite charge and heat current  contributions (in gray) that appear as prolongation of the main structures are effects of the finite coupling strength and they vanish in the limit of small coupling.

In \cref{fig5_IQ_vV} we plot the total occupation $N=n_1+n_2$, the charge and the heat currents as functions of the gate and bias at $\delta v=0$. In the upper left panel of \cref{fig5_IQ_vV}, the lines are solutions of the equations $p_{i,j}-V_{\alpha}=0$ where lines with positive (negative) slope correspond to $V_L$ ($V_R$).
Furthermore, the dashed lines correspond to poles $p_{1,j}$ related to site one while the solid lines correspond to poles $p_{2,j}$ related to site two. In the low
temperature and low coupling limit, these lines configure the regions of finite currents (bottom panels of \cref{fig5_IQ_vV}). The main features are three diamonds of vanishing currents which are determined by the intersection of the pole contributions $p_{1,1}=p_{2,1}$, $p_{1,5}=p_{2,5}$, $p_{2,3}$ and $p_{1,4}$. For a fixed bias voltage, a visual interpretation can be obtained if one takes the path  determined for fixed  $\delta v$ in the left panels of \cref{fig4_IQ_v1v2}. As the gates are varied , the transitions between the regions of integer local occupations  corresponding to the stripes are crossed. In particular, for $v_1=v_2$, the stripes transition regions are determined by the aforementioned poles $p_1$, $p_5$, $p_{2,3}$, and $p_{1,4}$.

\begin{figure}
\centering
\includegraphics[width=\linewidth]{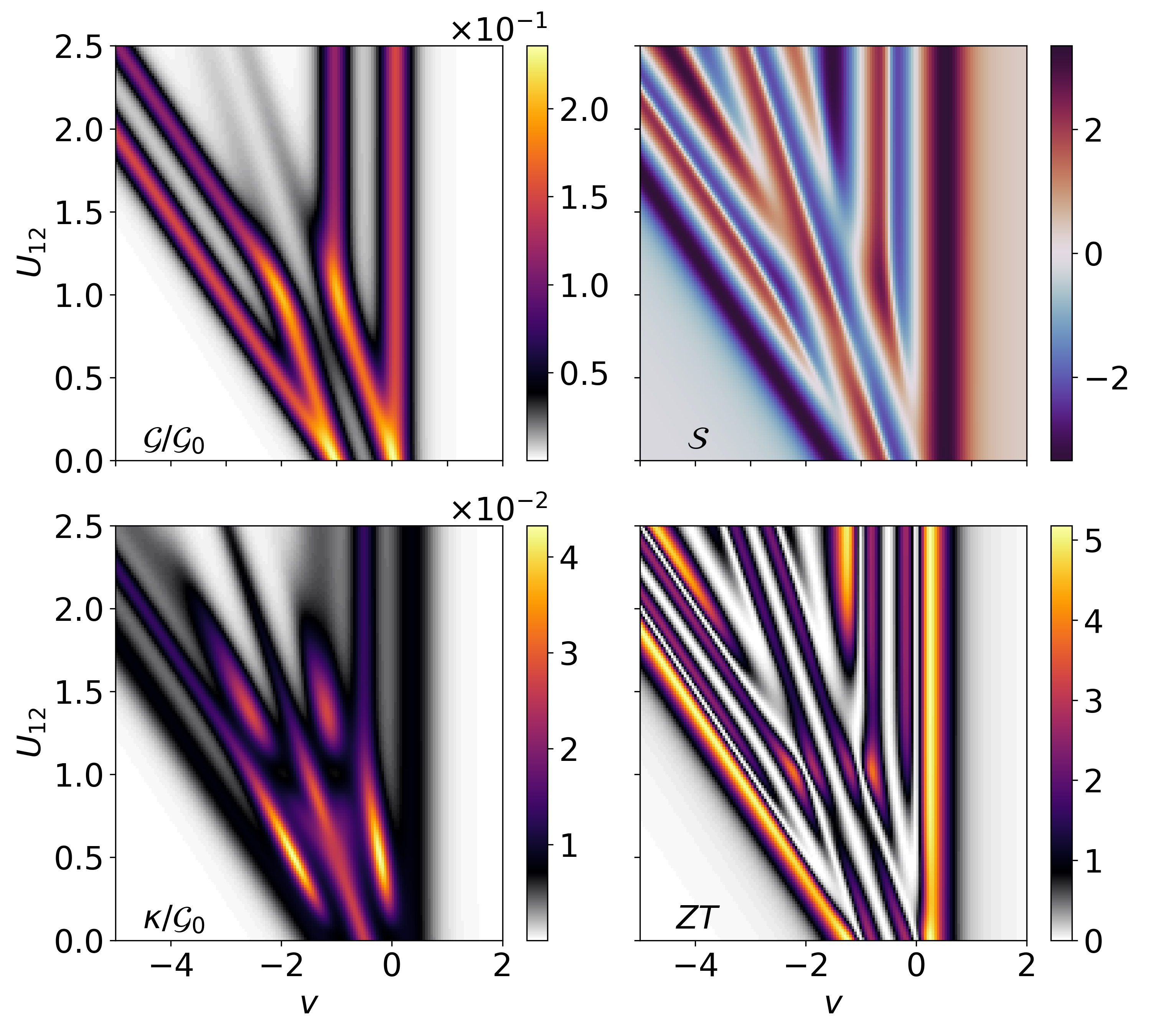}
\caption{Linear transport coefficients and figure of merit as a function of the gate level $v$ and the inter-Coulomb repulsion $U_{12}$ for fixed $T=2\gamma=0.1$. Energies in units of $U_1=U_2$.}
\label{fig6_LTC}
\end{figure}

  In \cref{fig6_LTC}, we present colormaps of the linear transport coefficients and the figure of merit as function of the common gate level $v=v_1=v_2$
  and the inter-dot Coulomb interaction $U_{12}$ at $U_1=U_2=1$, $T=0.1$ and $\gamma=0.05$. The overall structure of the transport coefficients is determined by the GF of the system. The conductance exhibits regions of high values determined by straight and diagonal lines in the plane corresponding to the resonance conditions where the energy levels of the DQD align with the Fermi levels of the leads. These resonances are determined by the poles of the GF, following the Coulomb blockade peaks at $p_{i,j}=0$. The two main poles that contribute at any
inter-dot Coulomb repulsion are $p_{i,1}$ and $p_{i,4}$. As the
inter-dot interaction is increased above $U_{12}>1$ the other finite contributions to the conductance peaks originate from the poles $p_{i,5}$ and $p_{i,3}$ to $p_{i,2}$ and $p_{i,6}$. Similarly, the Seebeck coefficient $\mathcal{S}$, displays a structure of vertical and diagonal diagonal bands where $S$ changes sign, indicating regions of strong thermoelectric response. These bands are aligned along specific directions in the $v$ and $U_{12}$ parameter space, again dictated by the poles of the GF. The highest values of $S$, both positive and negative, are observed at the intersections of the lines governed by $p_{i,1}$, $p_{i,4}$, $p_{i,2}$ and $p_{i,6}$. The electronic contribution to the  thermal conductance $\kappa_e$, shown in \cref{fig6_LTC}(c), also displays a structured pattern with peaks occurring at similar resonance conditions as the conductance and Seebeck coefficient. 
Interestingly, the highest $\kappa_e$ values are obtained for inter-Coulomb repulsion around $U_{12}\sim 0.5,1.5$. 

The figure of merit $\text{ZT}$, shown in \cref{fig6_LTC}(d), highlights regions where the system achieves optimal thermoelectric performance. High $\text{ZT}$ values are concentrated along the lines of resonance, particularly around $v = 0$, $v=-2U_{12}-U_i$, $v=-U_i$ and $v = -2U_{12}$. For example, at $v\approx 0.2$, $\text{ZT}$ reaches a maximum due the combined effects of high $\mathcal{S}$, moderate $\mathcal{G}$, and controlled $\kappa_e$. These regions signify the potential for efficient thermoelectric energy conversion in the DQD system.

\section{Conclusions}
\label{section_5}
In this paper, we have presented a fully analytical
solution of the equation of motions
for an interacting double quantum dot, each symetrically connected to two
  leads, in the Coulomb blockade regime out of thermal equilibrium.
The one-particle GF is obtained purely in terms of the local occupations and the interactions by solving a linear system for the density correlators analytically, following an equivalent derivation 
for thermal equilibrium.
Our out-of-equilibrium approach has the same formal structure as
  the equilibrium approach of Ref.~\cite{sobrino2024fully} with the only difference being that the relation \cref{eq_exp_values} between (out-of-equilibrium)
  density correlators and (higher-order) GFs now contains an average of two Fermi functions, each referring to one of the leads. This form alone incorporates all the dependencies on the driving forces and, compared to equilibrium, leads to a single modified definition for the function of \cref{eq_int_digamma}. 
Subsequently, the local occupations, the charge and heat currents, the transport coefficients, and the figure of merit can all be analytically expressed solely in terms of system parameters, interactions, and external driving forces (both bias and temperature gradient).

The evolution of the regions of stable occupations for finite bias is understood in terms of the poles of the one-particle GF of the system, which correspond to the addition and removal energies of the equilibrium situation, shifted due to the bias. The application of a finite bias  results in the formation of stripe regions of non-integer local occupations in the plane of the gate levels. These stripe regions exactly define the regions of non-vanishing charge and heat currents in the limit of low coupling and low temperature. The Coulomb blockade diamond structure of the currents in the gate-bias plane is also completely determined by the pole structure of the GF, allowing for an understanding of the far-from-equilibrium properties of the system under given working conditions.

The analytical results obtained with the EOM method were compared against numerical results obtained with the HEOM technique. Our approach correctly reproduces the emergence of the stripe regions as the bias is increased, accurately capturing the local occupations and the currents along different directions in the plane of the local gate levels and for various out-of-equilibrium configurations in the Coulomb blockade regime, thus validating the approximation of our derivation. Furthermore, our derivation of the linear transport coefficients allowed us to calculate the figure of merit, providing a comprehensive assessment of the thermoelectric performance of the DQD system and identifying regions with efficient thermoelectric energy conversion.

The analytical EOM approach provides significant advantages, including the ability to reveal explicit functional dependencies that elucidate the underlying physical mechanisms of the system. Additionally, this method is computationally efficient, enabling extensive exploration of different parameter regimes and the detailed study of transport properties.

\section*{Acknowledgements}

We acknowledge financial support by grant IT1453-22 “Grupos
Consolidados UPV/EHU  del Gobierno  Vasco” as well as through Grant
PID2020-112811GB-I00 funded by MCIN/AEI/\linebreak10.13039/501100011033.
We acknowledge technical support provided by SGIker (Scientific Computing
Services UPV/EHU).
N.S. acknowledges funding from the European Union under the Horizon Europe research and innovation programme (Marie Skłodowska-Curie grant agreement no. 101148213, EATTS). 
DJ acknowledges funding by the ``Plan Gen-T of Excellence'' of Generalitat Valenciana through grant CIDEXG/2023/7.
\bibliography{biblio}

\end{document}